\documentclass[traditabstract]{aa}
\usepackage{txfonts}
\usepackage{graphicx}

\begin{document}

\title{Sulphur in the metal poor globular cluster NGC~6397
  \thanks{This paper includes data gathered with the 6.5 meter Magellan Telescopes located at 
Las Campanas Observatory, Chile.}
}


\author{Andreas Koch,   \and Elisabetta Caffau}
\authorrunning{A. Koch  \& E. Caffau}
\titlerunning{Sulphur in NGC~6397}
\offprints{A. Koch \email{akoch@lsw.uni-heidelberg.de}}

\institute{Landessternwarte, Zentrum f\"ur Astronomie der Universit\"at Heidelberg, K\"onigstuhl 12, 69117 Heidelberg, Germany}
\date{}

\abstract {Sulphur (S) is a non-refractory $\alpha$-element that  is not
locked into dust grains in the interstellar medium. Thus no correction to the measured, interstellar sulphur
abundance  is needed  and it can be readily compared to the S content in stellar photospheres. 
Here we present the first measurement of sulphur in the metal  poor globular cluster (GC) NGC~6397, as detected in a  MIKE/Magellan high signal-to-noise, 
high-resolution spectrum of one red giant star.  While abundance ratios of sulphur are available for a larger number of Galactic stars down to an [Fe/H] of 
$\sim -3.5$ dex, no measurements in globular clusters more metal poor than $-$1.5 dex have been reported so far. 
We find a { NLTE, 3-D abundance ratio of [S/Fe] = +0.52} $\pm$0.20 (stat.) $\pm$0.08 (sys.), based on the { S\,I}, Multiplet~1 line at 9212.8\AA. 
This value is consistent with a Galactic halo plateau as typical of other $\alpha$-elements in GCs and field stars, but we cannot rule out 
its membership with a second branch of increasing [S/Fe] with decreasing [Fe/H], claimed in the literature, which leads to a large scatter at metallicities around $-2$ dex.
The [S/Mg] and [S/Ca] ratios in this star are compatible with a Solar value to within the (large) uncertainties. Despite the very 
large scatter in these ratios across Galactic stars between literature samples, this indicates that sulphur 
traces the chemical imprints of the other $\alpha$-elements in metal poor GCs. 
Combined with its moderate sodium abundance ([S/Na]$_{\rm NLTE}$=0.48), the [S/Fe] ratio in this GC extends a global, positive S-Na correlation that is not seen in field stars and 
might indicate that proton-capture reactions contributed to the production of sulphur in the (metal poor) early GC environments. 
}

\maketitle 
%
%
%
%
%
%
\section{Introduction}
The early chemical evolution of any stellar system is very efficiently traced by the chemical element distributions of the $\alpha$-elements (O, Ne, Mg, Si, S, Ar, Ca, and possibly Ti):  
these are produced in massive stars and dispersed into the interstellar medium (ISM) by supernovae (SNe) of type II 
on time scales much shorter than other reference elements like those of the iron group.  In particular,  the [$\alpha$/Fe] ratio has been adopted as a clear descriptor of 
the chemical evolutionary history of the Galaxy. 
While O, Mg, Si, Ca, and Ti are generally readily measurable in Galactic disk and halo stars, sulphur has always been the {\em enfant terrible} in stellar chemical 
abundance analyses due to the difficult measurability of its weak absorption lines, the strongest of which lie in the (near-) infrared and are often affected by contamination from telluric lines. 
On the other hand, as the volatile element S is not depleted onto dust it has been accurately determined in a number of damped Lyman $\alpha$-absorbers. This allows  for a
straightforward chemical tagging of the ISM in those gas-rich, early-type objects and therefore permits us to draw a chemical parallel with the early Galactic matter. 
 
Unfortunately, the findings regarding the [S/Fe] ratio in Galactic halo stars in the literature are ambiguous: 
while Israelian \& Rebolo (2001) and Takada-Hidai et al. (2002) report on a linear rise of this abundance ratio towards lower metallicities, below [Fe/H]$<-1$ dex, 
 this is not confirmed in the works by Ryde \& Lambert (2004),  Nissen et al. (2004, 2007), Spite et al. (2011), and J\"onsson et al. (2011).  
Those data rather indicate  a uniform enhancement to a plateau value of 0.35 dex after a strong increase between Solar metallicities down to $-1$ -- in accord with the canonical distribution of the $\alpha$-elements in the metal poor Galactic halo. 
Finally,  Caffau et al. (2005a) note the possibility of a bimodality in [S/Fe], branching at an [Fe/H] below $-1.1$ dex, or at least an increase in the scatter, where one fifth of their sample exhibits high values of [S/Fe] above the plateau. Likewise, Takeda \& Takada-Hidai (2011) note a complicated behavior, in which a local plateau contrasts a ``discontinuous jump'' of the S/Fe ratio at low metallicities.

A constant abundance with metallicity  is easy to reconcile with the notion that sulphur, as an $\alpha$-element, is produced like Si and Ca from O-burning during  the explosive SN phase with a hydrostatic contribution 
(Limongi \& Chieffi 2003).  
Thus it should also follow the same abundance trends like these elements (see also Fulbright et al. 2007; Koch \& McWilliam 2010, 2011; hereafter KM11). 
In contrast, high S/Fe ratios at low metallicities require alternative enrichment and mixing scenarios incorporating nucleosynthesis in massive ($\sim$100 M$_{\odot}$) hypernovae (e.g., Nakamura et al. 2001) or 
a fast and efficient mixing of volatile SNe ejecta into to the ISM relative to other (iron peak and refractory $\alpha$) elements (e.g., Ramaty et al. 2000).

Since Spite et al. (2011) remark  
that it is ``unclear whether S-rich stars [i.e., [S/Fe]$>$ +0.7] exist with metallicities in the interval $-2.5<$[Fe/H]$<-1.1$'', 
it appears crucial to bolster the Galactic sulphur-scale by detailed studies of its Globular clusters (GCs) as they pose important tracers of the earliest enrichment phases of the Galaxy. 
However, S-abundances have been investigated in only three GCs  to date (Caffau et al. 2005b; Sbordone et al. 2009), out of which the Sagittarius cluster Terzan 7 is not even a typical representative of the genuine 
Milky Way population. 
Furthermore, the observed light element variations in GCs (Gratton et al. 2004) and the presence of multiple stellar populations, likely coupled with variations in their $\alpha$-element content  (Lee et al. 2009), 
emphasize the need to explore 
whether any such inhomogeneities also hold for sulphur, as suggested for the metal rich GC 47~Tuc (Koch \& McWilliam 2008; Sbordone et al. 2009).

In this Letter we take a first step towards such studies by reporting on the first measurement of a sulphur abundance ratio in one star of the metal poor ([Fe I/H]=$-$2.1; KM11) GC NGC~6397, which is a close and well studied, archetypical Galactic halo GC (e.g., Richer et al. 2008; Lind et al. 2011; KM11, and references therein). 
 In \textsection 2 we present the data set and the derivation of the abundance ratio and its corrections, before discussing the results in the light of Galactic sulphur production in \textsection 3. 
\section{Data and Analysis}
The spectra for the red giant \#13414 were presented in KM11 and are based on  observations with 
the Magellan Inamori Kyocera Echelle (MIKE) spectrograph at the 6.5-m Magellan2/Clay Telescope, yielding very  high signal-to-noise (S/N) ratios. 
This target is the only of the sample to comprise the reddest echelle order containing suitable and/or uncontaminated S\,I-lines (Sect.~2.1).  

During the observing run in June 2005 also a few  telluric standards were  included, but none was exposed at  identical airmass to the GC star, prohibiting 
a perfect elimination of the copious telluric absorption features  (chiefly water vapor) in the near-infrared. 
However, the spectra, shown in Fig.~1,  are still suitable for measurements of the uncontaminated line at 9212.8\AA. 
\begin{figure}[!ht]
\begin{center}
\includegraphics[angle=0,width=1\hsize]{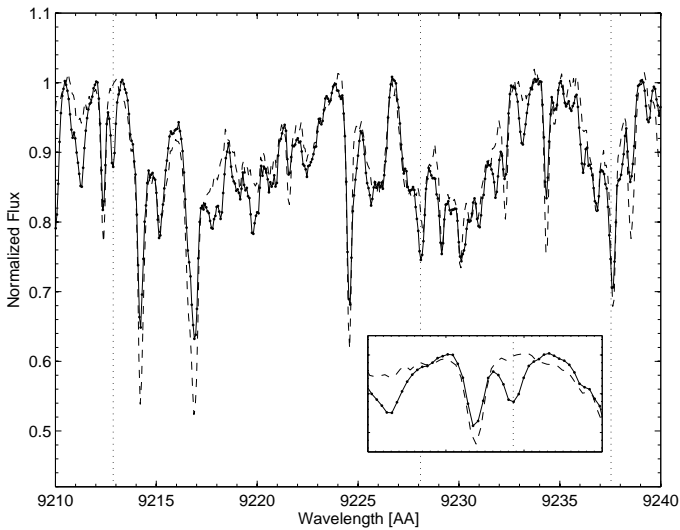}
\end{center}
\caption{Spectrum of star \#13414 (solid line) and the telluric standard HR~3090 (dashed line). Vertical lines indicate the rest wavelengths of Multiplet 1, while the inset highlights the region around the bluest of those lines at 9212.863\AA~that we used for EW measurements.}
\end{figure}

\subsection{Line data and measurements}
Here we derive a sulphur abundance based on the 9212.863\AA-line of the multiplet~1. 
Unfortunately, the giant's relative radial velocity of 14.6 km\,s$^{-1}$ shifts 
the other two multiplet lines at 9228.90, 9237.54\AA~ square on top of telluric lines, rendering the contamination too large to derive a meaningful 
result from those transitions.  
Likewise, all other multiplets commonly used in the analyses of solar metallicity or slightly metal-poor stars, such as Mult.~8 at 6750\AA~or Mult.~6 at 8694\AA, are too weak in the metal poor GC star  
and undetectable even in our high S/N ratio spectra, as verified by spectral synthesis.

In practice, we determined the sulphur abundance by measuring the equivalent width (EW) of the { S\,I}-line at 9212.8\AA; using a Gaussian fit in {\sc iraf}'s {\em splot} task yields an EW of 41 m\AA. At the strength of the feature and since this line is largely unaffected by the wing of the Paschen $\zeta$ line at 9229.0\AA, an EW analysis is expected to return reliable results. 
Although molecules have a non-negligible contribution to the opacities in cool stars like \#13414,  the 
S~{\sc i} line at 9212.8\AA~in our metal poor target is unaffected by adjacent (chiefly CN) molecular features, as we also verified by computing a synthetic 
spectrum with and without the sulphur line. 
For this particular transition, several oscillator strengths are found in the literature, spanning the range of log\,$gf$=0.38 to 0.47 (Lambert \& Warner 1968; Ryde \& Lambert 2004; see Table~1 in Caffau et al. 2005a). 
Here we adopt the value of  0.42 (Wiese et al. 1969) together with an excitation potential of 6.525 eV. 
At our observed EW of 41 m\AA, the exact choice of log\,$gf$, however, introduces a mere 1$\sigma$ scatter of 0.04 dex.

As in KM11, we assigned this star with a model atmosphere based on the Kurucz LTE atmosphere grid\footnote{{\tt http://kurucz.harvard.edu}}  without convective overshoot, using $\alpha$-enhanced opacity distributions  AODFNEW (Castelli \& Kurucz 2003)\footnote{{\tt http://wwwuser.oat.ts.astro.it/castelli}}. 
Furthermore, we used the stellar parameters derived by KM11 of  (T$_{\rm eff}$=4124 K, log\,$g$=0.29, $\xi$=1.74 km\,s$^{-1}$, [M/H]$\equiv$[Fe~I/H]=$-$2.14 dex) and performed 
our syntheses and abundance computations with the  the {\em synth} and {\em abfind} drivers of the 2010 version of the MOOG code (Sneden 1973). 

\subsection{Abundance errors and corrections}
The signal-to-noise (S/N) ratios around the S\,{I}-9212\AA~line in the red giant and the telluric standard, at  530 and 250 per pixel, 
are  very high and we can safely neglect any EW measurement error based on the noise component. Strong telluric absorption and the presence of many, albeit 
weak, molecular features in this near-infrared region renders continuum placement difficult across the entire order 
and thus provides the dominant contribution to the random uncertainty.  The narrow window around the actual line can be estimated reasonably well and 
a 1\% continuum uncertainty propagates into an EW error of $\pm$6 m\AA. Likewise, the adjacent telluric line at 9212.4\AA~is easily subtractable, 
leaving a residual contamination of no larger than 3 m\AA. 
Finally, we note the possible blending with a  weak Fe\,{I} line at 9212.97, the predicted EW of which, however, 
is not in excess of 1.2 m\AA~at the stellar parameters of \#13414. 
Adopting a conservative EW error of 10 m\AA~as a combination of these effects then translates into a 1$\sigma$ random error on [S/Fe] of 
0.20 dex. 

To quantify the systematic errors on our abundance ratio, we perform a 
standard error analysis in analogy to KM11, using the giant's stellar parameters and the variations thereof.
We thus find sensitivities to (T$_{\rm eff}\pm$50K, log\,$g\pm$0.2, $\xi$$\pm0.1$ km\,s$^{-1}$, [M/H]$\pm$0.1 dex, ODF\footnote{Uncertainty based on the 
use of $\alpha$-enhanced versus Solar-scaled opacity distributions; see KM11.}) of 
($\mp0.09$, $\pm0.08$, $\mp0.02$, $\pm0.02$, $-0.09$) dex, which we interpolate to the actual stellar parameter uncertainties set in KM11. 
Adding these in quadrature leads to an upper limit of the total, systematic uncertainty of 0.08 dex with the dominant contribution stemming from the effective temperature scale. 

{ In the following we discuss several corrections that need to be considered for an accurate description of our abundance results. 
A summary of these results and corrective terms is given in Table~1.}
\subsubsection{Non-LTE}
The formation of the S\,I multiplet lines is affected by departures from Local Thermodynamic Equilibrium (NLTE). 
While all the available grids of NLTE corrections in the literature as well as observations of sulphur in red giants only 
 extend to stars warmer  than $\sim$4500 K, those results suggest  that such effects in the cooler giants tend to be small (e.g., Takada-Hidai et al. 2002; 
Takeda et al. 2005; Spite et al. 2011). 
{ These results are, however,  dependent on the adopted cross section of hydrogen collisions, S$_H$. While the majority of abundance studies either fully neglects NLTE corrections or 
assumes a standard value of S$_H$=1, recent empirical studies lean towards lower efficiencies around $S_H$=0.01 (Mashonkina et al. 2011), which lead to significantly larger corrections (Table~1; 
in the sense of NLTE$-$LTE), both in S and Fe. 

The main objective of this work is to place NGC~6397 in the context of the Galactic chemical evolution of S and to address the role of a possibly bimodal distribution by comparison with 
all available literature sources. 
Unfortunately, a homogeneous sample is difficult to obtain, since different studies use different NLTE prescriptions (based on the used multiplets and stellar type) and not every work uses corrected Fe abundances, neither. 
Therefore, in order to maintain a decent level of homogeneity we show in Fig.~2 the 1D-LTE value {\em and} our corrected estimate, based on $S_H$=1 as found in the majority of literature S-abundance studies.}
\subsubsection{3-D}
The effects of 3-dimensional (3-D) atmosphere corrections were obtained by comparison with the data of Spite et al. (2011; in turn based on the calculations of 
Caffau et al. 2011a); these imply an upward correction to log\,$\varepsilon$(S) of 0.07 dex that is valid over a broader temperature range for giants. 
{ 

For the case of iron there are no suitable 3-D computations available in the literature. For an order of magnitude estimate, we consulted the models of Dobrovolskas et al. (2010), which are, however, 
only provided for a (T$_{\rm eff}$=5000K; log\,$g$=2.5)  atmosphere. Applying the corrections for the [M/H]=$-$2 case to the Fe~I line list of KM11 (as a function of excitation potential) 
we find  a 3-D correction, $\Delta_{\rm 3D}$=log\,$\varepsilon$(3-D)$-$log\,$\varepsilon$(1-D), 
 of $-0.05$ dex on average. However, the {\em mean corrected iron abundance ratio}, accounting for the random scatter in the line-by-line (differential) abundances, does not change upon the switch from 
 1-D to three-dimensional models. 
These line-by-line corrections will, however, induce changes of the excitation equilibrium and therefore cause changes in our adopted temperature scale. 

Both NLTE and 3-dimensional corrections, for S and Fe,  were ultimately 
simply added to our LTE, one-dimensional measurement. }
\subsubsection{Sphericity}
{ With a surface gravity of 0.29  our red giant has an extended atmosphere. Therefore, the spherical geometry of the model atmospheres is  expected to 
become important, in particular, since the S abundance is derived from a fairly high-excitation line ($\chi$=6.5 eV), while [Fe/H] is 
measured from lower-excitation Fe I lines (median $\chi$=4 eV) originating further out in the atmosphere. 

 We estimate the typical corrections to log\,$\varepsilon$(S) induced 
by the extended atmospheres based on computations\footnote{ These comprise OSMARCS model atmospheres (Gustafsson et al. 2008) 
 for T$_{\rm eff}$=4000 and 4250 K; log\,$g$=0.5; $\xi$=2 km\,s$^{-1}$; 
[M/H]=$-2$ dex  and employed the spectral analysis code ``Turbospectrum''  (Alvarez \& Plez 1998).}
 kindly provided to us by M. Spite. As a result, we find that our S-abundance based on  plane-parallel assumptions is lower than the value 
 based on a full, spherical treatment by no more than 0.1 dex (see also Drake et al. 1993).  

For the case of iron lines, we have to rely on the study of Heiter \& Eriksson (2006), who investigate differences between abundance analyses using plane-parallel versus spherical geometry atmosphere models. For the Fe\,I line list used in KM11, corrections are of the order of 0--0.25 dex for individual lines. 
We estimate the change in the {\em mean} [Fe/H], in a Monte Carlo approach on KM11's line list, to be no larger than +0.11 dex (in the sense of 
spherical$-$plane-parallel). We note, however, that the study of Heiter \& Eriksson (2006) only dealt with warmer (5000 K), extended (log\,$g$=0.5) stars, or cool stars (4000 K) with 
higher surface gravities  (log\,$g$=1) and no accurate corrections can be determined for the actual, cold, and extended star \#13414. 
Moreover, the value above is likely an upper limit, given the metal poor nature and thus small EWs of the Fe lines of KM11. 
Also note that the effects on S and Fe approximately cancel out so that the [S/Fe] abundance ratio remains unaffected by spherical geometry. 

Nonetheless, as above, the geometry of the atmospheres will upset excitation and ionization equilibria and thus affect our T$_{\rm eff}$ and log\,$g$ scale. 
On the other hand, KM11 derived their stellar parameters in a differential fashion and these effects can be considered negligible (see discussions in Koch \& McWilliam 2008). 
The respective influence on the systematic uncertainties of our measurements are beyond the scope of the present paper. }
\begin{table*}[htb]
\caption{Abundance results and corrections}             
\centering          
\begin{tabular}{ccccc|ccccc}     
\hline\hline       
log\,$\varepsilon(S)$   & \multicolumn{2}{c}{$\Delta_{\rm NLTE}$} &  & &
log\,$\varepsilon(Fe)$ & \multicolumn{2}{c}{$\Delta_{\rm NLTE}$} & \\
\cline{2-3}\cline{7-8}
1D, LTE & $S_H=1$ & $S_H=0$ &  \raisebox{1.5ex}[-1.5ex]{$\Delta_{\rm 3D}$} &  \raisebox{1.5ex}[-1.5ex]{$\Delta_{\rm spher.}$} & 
1D, LTE & $S_H=1$ & $S_H=0.01$ & \raisebox{1.5ex}[-1.5ex]{$\Delta_{\rm 3D}$} &  \raisebox{1.5ex}[-1.5ex]{$\Delta_{\rm spher.}$} \\
\hline
5.59$^a$ & $-$0.11$^b$ & $-$0.42$^b$ & +0.07$^c$ & +0.09$^d$ & 5.36$^e$ & +0.03$^f$ & +0.35$^f$ &$\sim$0.00$^g$ & $<$0.11$^h$\\
\hline                                    
\end{tabular}
\\{\em References:} $^a$This work; $^b$Takeda et al. (2005); $^c$Spite et al. (2011); $^d$M. Spite (private comm.); $^e$KM11; \\
$^f$Mashonkina et al. (2011); $^g$Dobrovolskas et al. (2010); $^h$Heiter \& Eriksson (2006).
\end{table*}
\section{[S/Fe] results}
Based on the above analysis we state here the  sulphur-to-iron abundance ratio of the red giant \#13414 in NGC~6397 as 
{ [S/Fe]$_{\rm NLTE, 3D}$ = 
+0.52$\pm$0.20(stat.)$\pm$0.08(sys.) dex, and 
 [S/Fe]$_{\rm LTE, 1D}$ = +0.59 dex.}
This assumes a solar abundance of log\,$\varepsilon_{\sun}$\,(S) = 
7.14 from the photospheric scale of  Lodders et al. (2009). 
Zero point shifts to other solar scales, such as the value of 7.16$\pm$0.05 employed by Caffau et al. (2011a) and Spite et al. (2011),  or the higher (LTE) value by Anders \& Grevesse 
(1989; 7.21) are small and will not affect our conclusions considering our conservative error bar. 
Iron- and all other element abundances for \#13414 are taken from KM11 in the following. 
Note that this star was characterized by KM11 as  a rather typical representative of this halo GC in that  its $\alpha$-elements (Mg, Si, Ca, Ti) are elevated to the canonical plateau value of [$\alpha$/Fe] $\sim$ +0.4 dex and that its light element  (O, Na, Mg, Al) distributions were affected by proton-capture reactions in an early generation of (massive AGB or fast rotating) stars (e.g., Ventura \& D'Antona 2008). 

In Fig.~2 we compare the measurement in NGC~6397 with the Galactic compilations of Caffau et al. (2005a)\footnote{This data set merged the author's own measurements and literature data from Israelian \& Rebolo (2001); Takada-Hidai et al. (2002); Chen et al. (2002, 2003); Ryde \& Lambert (2004); Nissen et al. (2004), and Ecuvillon et al. (2004).},  
Spite et al. (2011)\footnote{Note that we omit the C-rich EMP star CS~22949-037 ([Fe/H]=$-3.97$; [S/Fe]$<0.5$; McWilliam et al. 1995; Spite et al. 2011) from this and the following figures.}, Takeda \& Takada-Hidai (2011), and J\"onsson et al. (2011), 
bulge stars of Ryde et al. (2010), as well as the three only other 
reports on sulphur abundances in GCs in the literature: 47~Tuc and NGC~6752 (Sbordone et al. 2009) and the metal rich Sgr cluster Terzan~7 (Caffau et al. 2005b). 
It is important to note that all these different studies employed different S-lines and multiplets (see, e.g., the discussion in J\"onsson et al. 2011). 

\begin{figure}[!ht]
\begin{center}
\includegraphics[angle=0,width=1\hsize]{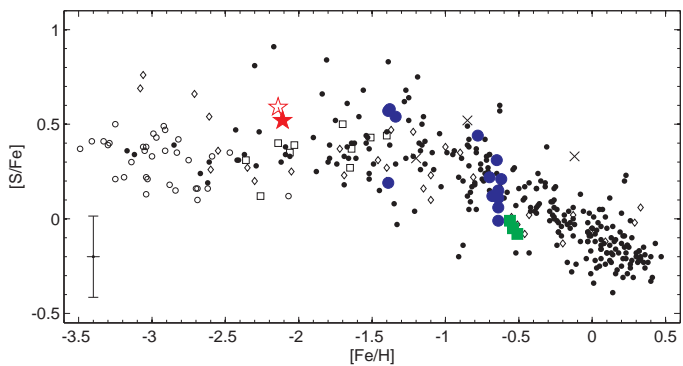}
\end{center}
\caption{[S/Fe] abundance ratios in galactic stars (black dots: Caffau et al. 2005a; crosses: Ryde et al. 2010; open squares: J\"onsson et al. 2011; open diamonds: Takeda \& Takada-Hida 2011), EMP stars (Spite et al. 2011, open circles), GCs from 
Sbordone et al. (2009, blue solid circles) and Caffau et al. (2005b, green solid squares), and { our measurement in NGC~6397 (red star), where the open symbol denotes the 1D-LTE value and the solid one is corrected for NLTE 
and 3-D. Our measurement error is indicated in the lower left corner.}}
\end{figure}

On the metal poor plateau, or, in the picture of a bimodal distribution, on the lower [S/Fe] branch of the distribution, the average S/Fe abundance ratio amounts to 0.35 dex (Ryde et al. 2004; 
Nissen et al. 2004, 2007; Caffau et al. 2005a; Spite et al. 2011). The compilation of Caffau et al. (2005a) contains 5 stars with abundances in excess of 0.7 dex that can be classified as ``S-rich''. Lastly, the highest value found in a 
GC star lies at 0.58 dex (Sbordone et al. 2009) -- fully compatible with our measurement in NGC~6397, albeit at a metallicity that is higher by 0.5 dex. 
Thus our observation of a metal poor GC star with a high S/Fe ratio that { skims the upper limit of the plateau} seems rather to indicate the {\em presence of a considerable abundance spread over a broad range in metallicity}. 
This is already seen in the interval of $-1.9 <$[Fe/H]$<-1.1$ dex and has been strengthened by the recent analysis of Takeda \& Takada-Hidai (2011), who find a discontinuous distribution, in which high-[S/Fe] stars were found below $-2.5$.  
On the other hand, at the lowest metallicities, the EMP stars both in the Spite et al. (2011) and Caffau et al. (2005a) samples only show little scatter in [S/Fe], 
with a 1$\sigma$ dispersion of $\sim$0.11 dex below an [Fe/H] of $-2.5$. 

Ryde \& Lambert (2004) suggested that the linearly rising trend seen in some studies is due to the use of the too weak Mult.~6 lines in the abundance analysis, coupled with 
uncertainties in establishing stellar metallicities from the Fe~I lines. 
Likewise, J\"onsson et al. (2011) advocate the use of [S I] line at 1082 nm and do not see evidence for any high-[S/Fe] stars. 
Since the present analysis is purely reliant on  Mult.~1 and our Fe~I abundance scale is based on a line-by-line differential analysis relative to a standard star (KM11), neither of these problems are expected to contribute and the moderately high value of [S/Fe]=0.67 we find is in accord with falling in between both branches, thereby contributing to the global scatter in [S/Fe]. 
\subsection{Correlations with light and other $\alpha$-elements}
One caution in the  comparison of sulphur with other elements in this GC star is that the results for Fe, Na, Mg, Ca and in KM11 were derived differentially with respect to the standard star Arcturus, while the present analysis of an S-abundance relies on, albeit well defined,  $gf$-values. In order to test the validity of such a comparison we 
also determined a differential [S/Fe] ratio for our star. To this end we measured the same S~I line in the Arcturus atlas of Hinkle et al. (2000) and placed \#13414 on the Arcturus abundance scale by using log\,$\varepsilon$(S) for Arcturus from Ryde et al. (2010) and its stellar parameters from Koch \& McWilliam (2008). 
The differential result is in excellent agreement  with the one cited above, using $gf$-values. 

In analogy to Si and Ca, sulphur  is chiefly produced by O-burning during the explosive SN II phase,  with similar contributions from the central burning or convective shell 
phases, as for Mg and O.   It is thus often suggested that these elements should trace each other closely with chemical evolution 
 (Kobayashi et al. 2006;  Fulbright et al. 2007; Koch \& McWilliam 2008, 2010, 2011). 
Therefore, we plot in Fig.~3 the [S/Mg, Ca] ratios as a function of the metallicity proxies [Fe, Mg, Ca/H] for the same Galactic components as introduced above\footnote{In order to 
obtain a fair comparison with the NLTE literature data, we applied the NLTE corrections for Na and Mg estimated in KM11 (in turn based on Takeda et al. 2003 and Andrievsky 
et al. 2010).}. 
\begin{figure}[!ht]
\begin{center}
\includegraphics[angle=0,width=1\hsize]{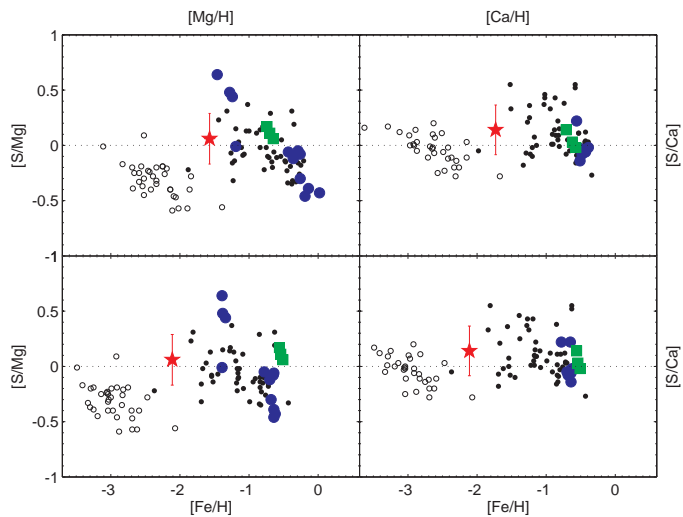}
\end{center}
\caption{[$\alpha_1/\alpha_2$] element ratios in the halo and GC stars shown in Fig.~2. 
The symbols are the same as above, with abundances for Ter~7 from Sbordone et al. (2007).  The dotted line indicates the Solar level at zero dex.}
\end{figure}

Caffau et al. (2005a) detect a large scatter in their S/Mg abundance ratios\footnote{Note that those authors employed the {\em LTE} Mg-abundances of Gratton et al. (2004).}, 
with a possible hint at an increasing ratio towards lower metallicities. 
While the mean element ratio in their sample lies around Solar, the mean [S/Mg] of $-0.32 \pm 0.14$ dex  in Spite's et al. (2011) EMP stars
is significantly lower and only shows small scatter. 
Intriguingly, also the latter authors suggest a possible rise of S/Mg with decreasing [Fe/H]; it is clear that, upon combining both samples, no such trend pertains. 
On the other hand, the GC data of Sbordone et al. (2009) show a very strong linear increase, which is dominated by the three high [S/Mg] stars in NGC~6752. 
We emphasize, however, that the Mg abundances used by Sbordone et al. were based on 1D-LTE calculations. 
Our value for S/Mg in NGC~6397 is fully  compatible with zero to within the uncertainties and therefore agrees with 
the bulk of the more metal rich sample above [Fe/H]$\ga -2$, suggesting that sulphur and magnesium indeed vary in lockstep 
safe for the most metal poor regime. 

The situation for the S/Ca ratio seems much simpler. Despite a small systematic offset between the EMP and halo star samples of 0.18$\pm$0.30, this abundance ratio is consistent with the notion of a common production of S and Ca, following the chemical evolution of the halo {\em and} its GCs.   

In Fig.~4 we investigate the connection of sulphur with the light element sodium. 
\begin{figure}[!ht]
\begin{center}
\includegraphics[angle=0,width=1\hsize]{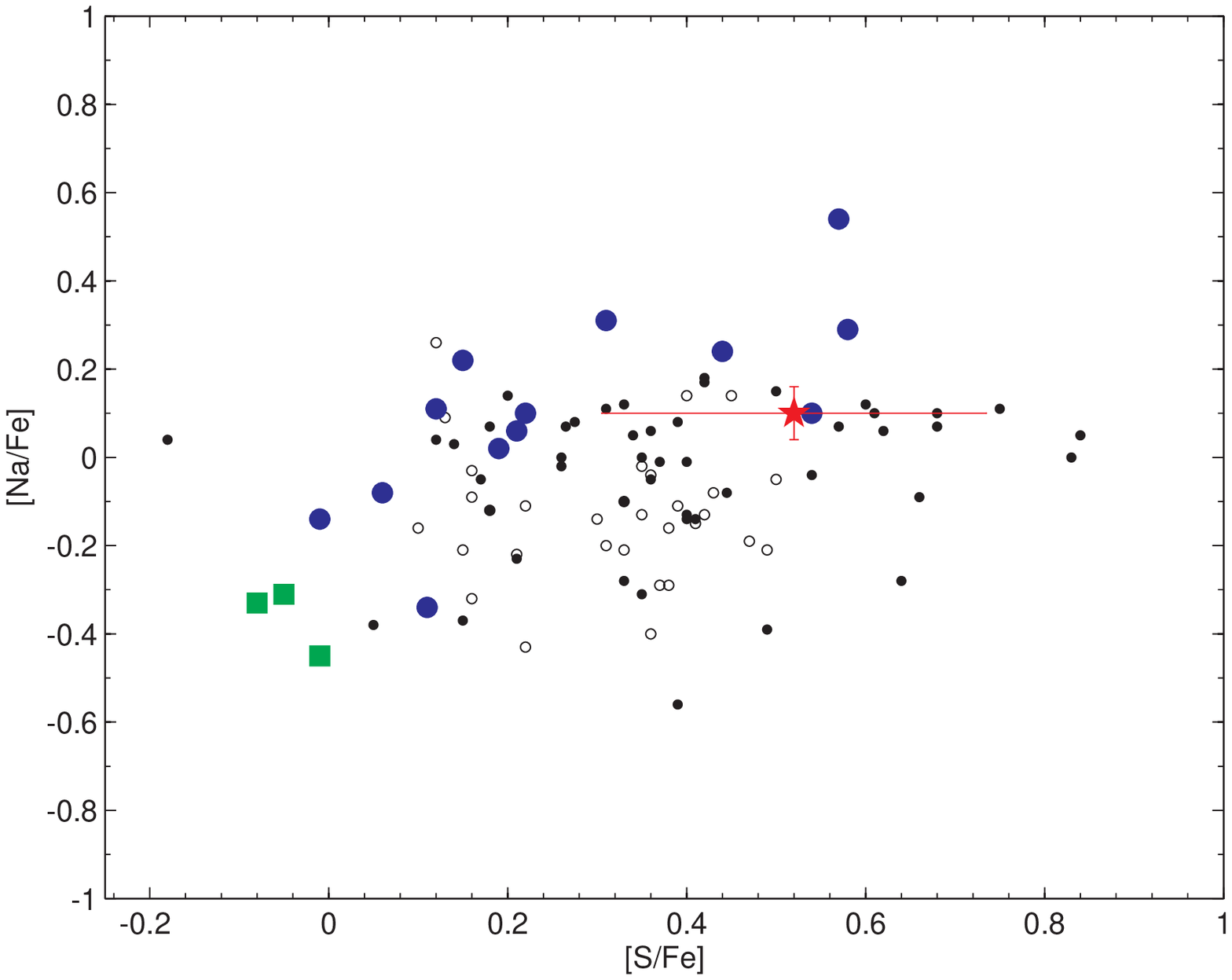}
\includegraphics[angle=0,width=1\hsize]{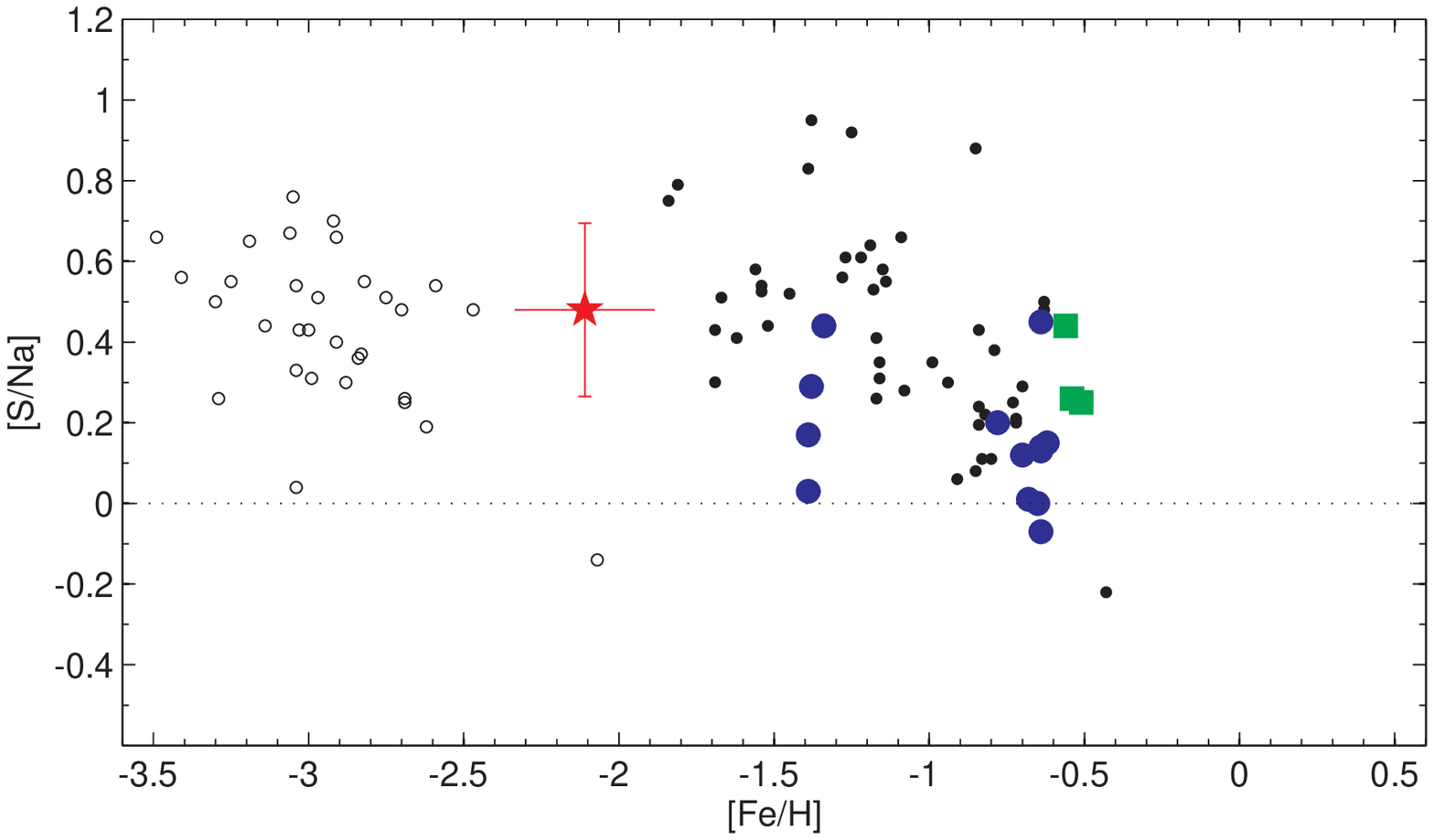}
\end{center}
\caption{S/Na correlation. Symbols are the same as in Figs.~2,3.}
\end{figure}
While the large [S/Fe] value is compatible with the larger LTE value of Na (and also Si) in this star (KM11), 
the estimated [S/Na]$_{\rm NLTE}$ of { 0.48} is fully in accord with those found in Galactic halo stars over the full metallicity range.  
On the other hand, this value is the largest found in a GC to date and emphasizes the possibility of a trend of increasing [S/Na] with decreasing [Fe/H] (see bottom panel of Fig.~4). 
In fact, Sbordone et al. (2009) note that, while there is no evidence of any  scatter in their GCs' [S/Fe] to within the (large) uncertainties, 
the presence of a  significant S-Na correlation clearly suggests that sulphur inhomogeneities do exist in the GCs. 
We note that all of NGC~6752, 47~Tuc, {\em and} NGC~6397 exhibit  the canonical light-element variations  in terms of 
Na-O, Mg-Al, and Na-Li (anti-) correlations (Carretta et al. 2009; Lind et al. 2009). 

In practice, the Pearson correlation coefficients of the run of [Na/Fe] vs. [S/Fe] in the combined samples (accounting for a global errorbar of 0.2 and 0.07 dex on S and Na, respectively; e.g., Fig.~2 in Sbordone et al. 2009) are 
0.58$\pm$0.13 (GCs), 
0.12$\pm$0.11 (field stars; Caffau et al. 2005a), and 
0.12$\pm$0.09 (combined field and EMP stars of Caffau et al. and Spite et al. 2011). 
Likewise, the correlations in [S/Na] vs. [Fe/H] space are 
$-$0.43$\pm0.09/$ (halo), 
$-$0.01$\pm0.07$ (halo and EMP), and
$-$0.20$\pm0.20$ (GCs).  
Thus the halo field stars appear not to exhibit any such correlations, in analogy to the Na-O relation, 
while abundance variations between sulphur and sodium are present amongst the GC sample; we cannot rule out that NGC~6397 was governed by the reactions producing these imprints.  

As elaborated in Sbordone et al. (2009), a positive correlation between S and Na is difficult to reconcile with the classical pollution and enrichment scenarios of GCs that 
produce, amongst others, the Na-O {\em anti-}correlation. 
In particular,  it is far more difficult to obtain low-S, low-Na stars in such a scenario, 
whereas the moderately high [Na/Fe] and the elevated [S/Fe]  found in our NGC~6397 red giant are consistent with  
being part of the second (''intermediate''; Carretta et al. 2009) cluster population that was polluted by material that underwent proton-capture reactions. 

Sbordone et al. (2009) suggest proton-capture as a significant source of sodium and sulphur. 
Indeed the very high abundance of phosphorus, at [P/Fe]$\sim$+3.1, [P/S]$\sim$+2.5, 
respectively, observed in horizontal branch stars in NGC~6397 (Hubrig et al. 2009) 
is believed to provide sufficient seed material to guarantee an ample enhancement 
in sulphur through the $^{31}$P(p,$\gamma$)$^{32}$S reaction  
without violating any of the observed (anti-)correlations (e.g., with Si) resulting from the possible other p-chains. 
One should however be cautious since the large overabundance of phosphorous in
Horizontal Branch stars could, at least partly, result from atomic
diffusion in the atmospheres of these stars (Michaud et al. 2008).
Furthermore, at higher metallicities, among field stars Caffau et al. (2011b) 
find a constant [P/S] value. Thus, if confirmed, an intrinsic high ratio 
[P/S] in NGC~6397, is probably due to a chemical evolution of the GC
different from that of field stars.
\section{Summary \& Conclusion}
We determined the chemical element abundance of sulphur in the metal poor GC NGC~6397. 
Although our sample consists of only one single star, measurements of [S/Fe] in Galactic GCs are sparse, with only 
16 red giant data reported in the literature to date. 

The [S/Fe] ration we find for this star is { elevated and}   
marginally consistent with the metal poor ``plateau'' of Spite et al. (2011).
Its iron abundance puts it { within}  the region of the large scatter
of Caffau et al. (2005a).
Two of the most metal-poor stars of Sbordone et al.  (2009) show a similar [S/Fe] ratio.
With this result we cannot distinguish, yet, between a well defined plateau 
in the intermediate metallicity range ($-2.5$$\le$[Fe/H]$\le$$ -1$) plus a bifurcation towards high S/Fe ratios, or if the sulphur production channels in this interval 
merely entail a very large scatter.   
Given the global scatter, further investigations of this recalcitrant chemical element in GCs are necessary to state if the 
[S/Fe] ratios in GCs are, as it seems, largely different from those in Galactic halo field stars. 
The same holds for the [S/$\alpha$] ratios that indicate that sulphur traces the same chemical signatures of chemical evolution as the other $\alpha$-elements (Mg, Ca) -- 
in field stars and in GCs. 

\begin{acknowledgements}
We are very grateful to Monique Spite for her help with spherical models and to Andrew McWilliam for providing the spectra for this star.  
The referee, Andreas J. Korn, is thanked for a valuable report that helped to improve the NLTE and 3D discussions of our results. 
AK acknowledges the 
Deutsche Forschungsgemeinschaft for funding from  Emmy-Noether grant  Ko 4161/1. 
\end{acknowledgements}
\end{document}